\documentclass[11pt,a4paper]{article}

\usepackage[percent]{overpic}
\usepackage{a4wide}
\usepackage[utf8]{inputenc}
\usepackage{multirow}	
\usepackage[export]{adjustbox}[2011/08/13]
\usepackage{mathrsfs}
\usepackage{tikz}
\usepackage{url}
\usepackage{slashed}

\usepackage{graphicx}
\usepackage{dcolumn}
\usepackage{bm}

\usepackage[english]{babel}

\usepackage{cite}

\usepackage{graphicx}
\usepackage{amsmath}
\usepackage{caption}
\usepackage{subcaption}
\usepackage{lmodern}

\usepackage{siunitx}

\usepackage{nicefrac}
\usepackage{xfrac}

\usepackage{graphicx}
\usepackage{caption}
\usepackage{subcaption}
\makeatletter
\def\endfmffile{%
	\fmfcmd{\p@rcent\space the end.^^J%
			end.^^J%
			endinput;}%
	\if@fmfio
		\immediate\closeout\@outfmf
	\fi
	\ifnum\pdfshellescape=\@ne
		\immediate\write18{mpost \thefmffile}%
	\fi}

\usepackage{slashed}

\newcommand{\shat}{\hat{s}}
\newcommand{\that}{\hat{t}}

\newcommand{\tdm}[1]{\mbox{\boldmath $#1$}}  
\newcommand{\be}{\begin{equation}}
\newcommand{\ee}{\end{equation}}

\usepackage{microtype}

\begin{document}

\title{\textbf{Hard color singlet BFKL exchange and \\gaps between jets at the LHC\footnote{Extended version of a presentation at the Forward and Small-x QCD workshop at CERN, January 2017.}}}
\author{\textbf{Andreas Ekstedt, Rikard Enberg and Gunnar Ingelman}\\[3mm]
Department of Physics and Astronomy, Uppsala University,\\ 
             Box 516, SE--751 20 Uppsala, Sweden}

\maketitle

\begin{abstract}
\vspace{0.5cm}
We explore the perturbative QCD dynamics of hard parton-parton scattering through the exchange of a color singlet two-gluon ladder as described by the BFKL equation, resulting in a rapidity gap between two high transverse momentum jets. Implementing this in a complete Monte Carlo event simulation that also accounts for additional QCD processes at softer scales provides dynamical modeling of gap survival probabilities, which makes possible a detailed comparison with data on such jet-gap-jet events. New data from CMS at the LHC extend the dynamic range of the previous Tevatron data, and can be reproduced reasonably well provided that the Soft Color Interaction model is modified based on the idea of reduced resolution power of softer gluon exchanges. This indicates the need for further theoretical developments in connection with other color exchange processes related to rapidity gaps in the hadronic final state.  
\end{abstract}

\section{Introduction}

\begin{figure}[ht]
\centering
   \includegraphics[width=0.3\linewidth,center]{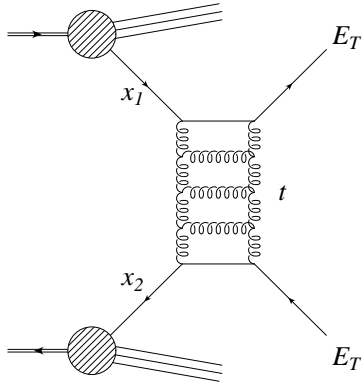}
\caption[asd]{Hard parton-parton scattering via a color singlet two-gluon ladder described by the BFKL equation.}
\label{fig:gluonladder}
\end{figure}

The idea of observing events in hadron-hadron collision with a large rapidity gap centrally between two high-$p_T$ jets was introduced by Mueller and Tang (MT)~\cite{Mueller:1992pe}, and was subsequently discovered experimentally at the Tevatron by the CDF and D0 collaborations \cite{Abe:1997ie,Abbott:1998jb}. Mueller and Tang calculated the cross section for elastic scattering of partons from exchange of a gluon ladder in a color singlet state at substantial momentum transfer (see Fig.~\ref{fig:gluonladder}), as described by the Balitsky-Fadin-Kuraev-Lipatov (BFKL) dynamics of QCD \cite{BFKL}. The gaps between jets signal has later been considered by several groups \cite{Cox:1999dw,Enberg:2001ev,Motyka:2001zh,Chevallier:2009cu,Kepka:2010hu}, and even though the original MT calculation, which only keeps the asymptotic rapidity dependence, does not give a good description of the data, a more refined BFKL calculation with non-asymptotic terms does give a good description of the rapidity and transverse energy dependence of the events at the Tevatron~\cite{Enberg:2001ev}.

However, in addition to the hard process in the form of BFKL exchange, it is necessary to take proper care to describe additional parton emissions and soft QCD effects, since any extra activity in the event is likely to destroy the gap that would otherwise have been produced. The simplest alternative is to compute the cross section and to multiply with a phenomenological gap survival factor, but we instead take these effects into account by performing a full Monte Carlo simulation using the event generator Pythia in conjunction with the LHAPDF6 library~\cite{Sjostrand:2006za,Buckley:2014ana}. In addition to the model for multiple parton-parton interactions present in Pythia, we used the Soft Color Interaction (SCI) model~\cite{Edin:1995gi,Edin:1996mw} for color rearrangements in the final state through soft gluon exchanges, since such rearrangements can have large effects on rapidity gaps.

In this note we extend the calculation~\cite{Enberg:2001ev} to LHC energies, and compare the results with the recent 7 TeV CMS data on gaps between jets~\cite{CMSdata}. The parton level cross section for color singlet elastic scattering of partons is obtained from a solution of the non-forward BFKL equation, as described in Section~\ref{sec:BFKL}. This solution is implemented as a hard process in the Monte Carlo event generator Pythia 6.4~\cite{Sjostrand:2006za}, which adds initial and final state parton showers, multi-parton interactions, and hadronization through the Lund string model~\cite{Andersson:1983ia} to account for the underlying event. In order to appropriately model the gap destruction processes, we also implement in the Pythia simulations both the original SCI model and a further development of this model. The treatment of these QCD effects and the simulations will be described in Section~\ref{sec:QCD}, followed by a comparison of model results with data in Section~\ref{sec:Results} and a concluding discussion in Section~\ref{sec:Conclusions}.

\section{Non-forward BFKL as origin of gaps between jets}
\label{sec:BFKL}

We calculate elastic parton-parton scattering of partons in the proton using the non-forward BFKL equation \cite{BFKL}. As shown in \cite{Mueller:1992pe,Bartels:1995rn} and discussed in \cite{Enberg:2001ev}, at large momentum transfer, the BFKL pomeron can be considered a hard probe of the proton structure, and predominantly couples to one single parton from the proton. We may therefore compute the differential cross section for elastic parton-parton scattering  
\be
\frac{d\hat\sigma (\shat,\that)}{d\that} = \frac{1}{16\pi} |A(\shat,\that)|^2,
\label{eq:dsigdt}
\ee
where $A(\shat,\that)$ is the amplitude and hats denote quantities on the parton level. We may then convolute this with normal collinear parton distribution functions (PDF) to obtain hadron-level cross sections. We will for concreteness first consider quark-quark scattering, but will later include also gluons.

If we define $\shat$ and $\that$ as the center-of-mass energy squared and momentum transfer in the partonic system, respectively, then in the high-energy limit $\shat \gg -\that \gg \Lambda_\text{QCD}$, the momentum transfer is dominated by the transverse momentum, and the amplitude is dominated by its imaginary part. In the BFKL framework, this amplitude for elastic scattering via BFKL pomeron exchange is given by a convolution of the BFKL Green's function that describes the exchange of the BFKL pomeron ladder with the \emph{impact factors} that describe the coupling of the pomeron to the scattered objects. This can be written in the form \cite{Lipatov:1996ts}
\begin{equation}
\mathrm{Im}\, A(\hat{s},\that) = \int \frac{d^2\tdm k}{(2\pi)^2} 
\frac{\Phi^{ab}_0(\tdm k, \tdm Q) \Phi^{ab}(x,\tdm k,\tdm Q)}{[(\tdm Q /2 + \tdm k) ^2 +s_0][(\tdm Q/2 - \tdm k)^2 +s_0]},
\label{eq:ampl}
\end{equation}
where $\Phi^{ab}_0(\tdm k, \tdm Q)$ is the impact factor for quark-quark scattering and $\Phi^{ab}(x,\tdm k,\tdm Q)$ is the BFKL-evolved impact factor.\footnote{If the function $\Phi^{ab}(x,\tdm k,\tdm Q)$ is replaced in Eq.~(\ref{eq:ampl}) by the unevolved impact factor $\Phi^{ab}_0(\tdm k, \tdm Q)$, one obtains the amplitude for two-gluon exchange.} Here we have defined $x= |\that|/\shat$ and $\that=-Q^2$, with $\tdm Q$ the transverse momentum of the BFKL pomeron. The momentum $\tdm k$ is defined so that the transverse momenta 
of the two exchanged gluons are $\tdm Q /2 \pm \tdm k$. We have also included an energy scale parameter $s_0$, which will be discussed below.

The impact factor $\Phi^{ab}_0(\tdm k, \tdm Q)={(\delta^{ab}/ 2N_c)} \Phi_0(\tdm k, \tdm Q)$ 
describes the coupling of the quark to two gluons with adjoint color indices $a,b$ 
in the color singlet state, jointly carrying the transverse momentum $\tdm Q$. We will also consider quark-gluon and gluon-gluon scattering, which may be obtained from the quark-quark impact factor by including the appropriate color factors. In the final amplitude this will give an overall color factor that depends on the scattered partons, as described below.

We go beyond the strict leading logarithmic approximation and take the running of the QCD coupling $\alpha_s = g^2_s / 4\pi$ into account. Then the impact factor can be taken as 
$\Phi_0(\tdm k, \tdm Q)= g_s ((\tdm Q /2+\tdm k)^2 +s_0) g_s((\tdm Q /2-\tdm k)^2+s_0)$.
$\Phi^{ab}$ is the result of the BFKL $x$-evolution from $\Phi_0 ^{ab}$
and is decomposed as
$\Phi^{ab} (x,\tdm k,\tdm Q) = {(\delta^{ab} /2N_c)}\Phi(x,\tdm k,\tdm Q)$.

The function $\Phi(x,\tdm k,\tdm Q)$ satisfies the non-forward BFKL equation \cite{BFKL}, which has an analytic solution due to Lipatov~\cite{Lipatov:1985uk}. Lipatov's solution is valid for the scattering of colorless particles, but Mueller and Tang (MT) computed the amplitude for quark-quark scattering using this solution \cite{Mueller:1992pe}, by making a modification to take into account colored particles. They moreover made the approximation of keeping only the leading \emph{conformal spin}\footnote{See e.g.~\protect\cite{Motyka:2001zh,Enberg:2002zy} for discussions.}, which is valid for asymptotically large rapidities, but it was later shown that this is not a good approximation for the moderate rapidities encountered in experiments~\cite{Enberg:2001ev,Motyka:2001zh}. This has also been seen for large-$t$ vector meson production at HERA~\cite{Enberg:2002zy,Enberg:2003jw,Poludniowski:2003yk,Enberg:2004rz}.

We avoid this issue by using the numerical solution of the non-forward BFKL equation obtained in \cite{Enberg:2001ev}, based on the method from \cite{Kwiecinski:1998sa}. This further allows us to include an approximate treatment of some of the next-to-leading (NLL) logarithmic corrections to the BFKL evolution. These are, first, the use of a running strong coupling $\alpha_s$, both in the impact factors and in the BFKL equation, and second, a restriction on the momenta in the real emission term in the kernel \cite{CC}. These are both formally subleading corrections to the leading logarithmic BFKL equation, but in practice they can be numerically large. The full NLL BFKL kernel is known~\cite{NLLBFKL}, but has a complicated form and its use is beyond the scope of this paper. 

The momentum restriction corresponds to a resummation of collinear divergences in the NLL BFKL kernel, and it has been demonstrated that for the forward BFKL kernel, the equivalent restriction accounts for about 70\% of the NLL corrections to the predicted pomeron intercept \cite{CC}. The effect is expected to be somewhat lower for non-forward BFKL~\cite{Enberg:2002zy}. For the running coupling we use the one-loop four-flavor $\alpha_s$ with $\Lambda_\text{QCD}=200$~MeV and scale $\mu^2=k^2+Q^2/4 + s_0$. 

Let us now finally discuss the parameter $s_0$. This parameter is related to the cancellation of infrared divergences in the amplitude, which is infrared finite, while the individual contributing diagrams is not. But our interpretation is that $s_0$ is related to confinement and parametrizes the fact that gluons do not propagate over distances much longer than $1/s_0$~\cite{s0}. In \cite{Enberg:2001ev}, $s_0$ was varied in the range $0.5 \text{ GeV}^2 < s_0 < 2 \text{ GeV}^2$, but here we use the fixed value $s_0=1$~GeV$^2$, which gave a good fit to Tevatron data. 

The non-forward BFKL equation has also been used to explain another process with rapidity gaps, namely large-$t$ diffractive photoproduction of vector mesons at HERA, which gives a good description of data~\cite{Bartels:1995rn,Bartels:1996fs,Forshaw:2001pf,Enberg:2002zy,Enberg:2003jw,Poludniowski:2003yk} (see \cite{Enberg:2004rz} for a review), and it may be argued that these two large-$t$ processes are the best places to look for BFKL dynamics.

\section{Additional QCD effects and gap survival}
\label{sec:QCD}

\begin{figure}[bth]
\centering
   \includegraphics[width=0.45\linewidth,center]{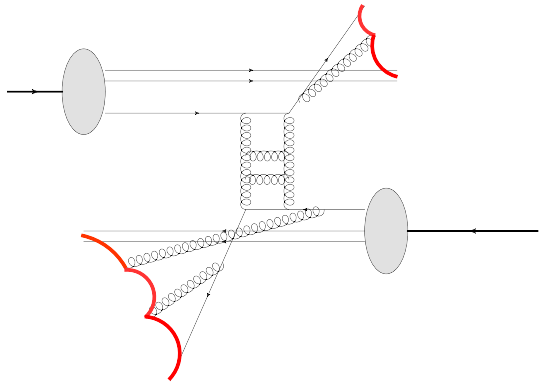}
\caption[asd]{Hard parton-parton scattering via the two-gluon BFKL ladder and additional initial and final state pQCD parton emissions giving the basic color structure with two color singlet string-fields separated by a central rapidity gap.}
\label{fig:ProtonColString}
\end{figure}

Based on the above cross-section for the perturbative process of two-gluon color singlet ladder exchange one expects a substantial rate of events with two jets with a rapidity gap in-between as illustrated in Fig.~\ref{fig:ProtonColString}. The Rutherford-like hard scattering via the dominating $t$-channel BFKL ladder implies that the two scattered partons mostly emerge in the same hemisphere as their respective proton remnant. Each of these parton-remnant systems are color singlets as the 2-gluon BFKL exchange carries no net color charge. Moreover, this basic color structure of the event is essentially unchanged by initial and final state parton showers which are dominated by collinear radiation emerging in a rapidity range given by the remnant and the hard-scattered parton. Thus, within the two color singlet systems separate color string-fields are formed, which hadronize to produce hadrons in their respective rapidity regions. This leaves a central rapidity region without any hadrons---i.e., a rapidity gap. 

However, with additional interactions there may be color exchanges resulting in additional color string systems or modified color string topologies. Strings may then be formed across the central rapidity region and their subsequent hadronization produces hadrons in the gap region. Thus, one must also take into account any additional activity in an event. Of particular importance is here multiple interactions (MI), i.e.\ additional parton-parton scatterings treated with conventional pQCD $2\to 2$ matrix elements. One or more such scatterings may occur in the same proton-proton collision, but ordered in their momentum transfers, which are constrained to be below the primary BFKL exchange $\sqrt{\hat{t}}$ and above a cut-off $p_{\perp 0}$ with a value of $1.5-2$ GeV as extracted from underlying event properties in normal jet events.  

\begin{figure}[bt]
\centering
   \includegraphics[width=0.7\linewidth,height=8cm,center]{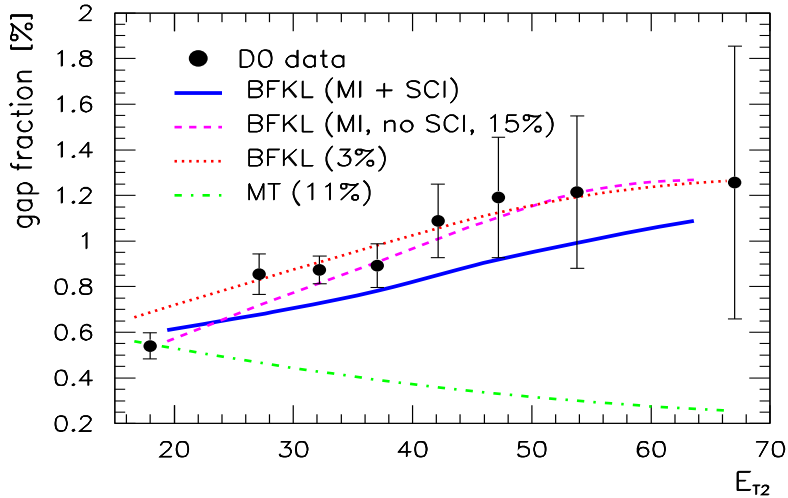}
\caption[asd]{Fraction (in \%) of 2-jet events having a rapidity gap in $|\eta|<1$ between the jets, versus the second highest jet-$E_T$. D0 data \cite{Abbott:1998jb} compared to BFKL color singlet exchange and with underlying event treated in different ways: simple 3\% gap survival probability, multiple interactions (MI) and hadronization requiring a 15\% gap survival probability, MI plus soft color interactions (SCI) with no need for an overall normalization factor. Also shown is the Muller-Tang calculation with an 11\% gap survival probability. From \cite{Enberg:2001ev}. }
\label{fig:Tevatron-ET}
\end{figure}

Taking these pQCD effects into account together with conventional Lund string hadronization in Monte Carlo event simulation, giving also substantial event-by-event fluctuations of all these processes, destroys a substantial fraction of the gaps between jets as compared to the parton-level BFKL result. Still, the remaining fraction of jet-gap-jet events over all dijet events was in our previous study \cite{Enberg:2001ev} found to be much larger than observed in Tevatron data. A simple way out is to introduce a multiplicative gap survival factor $S^2$ as a constant value and we obtained 15\% from fitting the Tevatron data. Such a factor is, of course, just a simple model for all additional gap destroying processes that are not accounted for. The Tevatron data could, however, be well described without such a factor, but instead including the Soft Color Interaction (SCI) model developed for other non-perturbative QCD processes and having only one free parameter already fixed by those. This comparison of models with the Tevatron data on jet-gap-jet events is shown in Fig.~\ref{fig:Tevatron-ET}. 

The Soft Color Interaction model was originally developed in order to explain rapidity gap events in hard diffraction at HERA~\cite{Edin:1995gi,Edin:1996mw} and was later used to describe hard diffraction at hadron colliders~\cite{Enberg:2001vq,Enberg:2002id,Enberg:2002tw,Ingelman:2012wj} as well as color rearrangements in quarkonium production~\cite{BrennerMariotto:2001sv} and B-meson decays~\cite{Eriksson:2008tm}. In all these cases the SCI model successfully describes the available data with only one free parameter, which specifies the probability for the exchange of a soft color octet gluon between any pair of partons in the partonic final state emerging from the underlying hard pQCD processes. 

Later theoretical developments \cite{Brodsky:2004hi,Pasechnik:2010cm,Pasechnik:2010zs} have provided a better theoretical understanding of such soft color interactions through explicit calculations of multiple gluon exchange between the hard-scattering system and the proton remnant in deep inelastic lepton-nucleon scattering. This has recently~\cite{Ingelman:2015qrt} been shown to give a good description of diffractive rapidity gap data from HERA.  

The original SCI model is formulated in terms of the emerging color charges from a hard scattering having a fixed probability to exchange color with one another. An essential implication is that the probability of color exchange depends on the parton multiplicity due to combinatorial effects. The parton multiplicity increases with the overall collision energy and the transverse energy scale of the hard interaction due to increased phase space for multiple interactions and parton radiation. It is, therefore, expected that the effective gap survival probability decreases when going from Tevatron to LHC energies, as is also observed in the new jet-gap-jet data of CMS~\cite{CMSdata}.

We find, however, that the original SCI model produces a too large gap destruction at LHC energy and attribute this to the increased parton multiplicity. In accordance with the developments in \cite{Ingelman:2015qrt}, we argue that the color exchange via soft gluons cannot resolve each individual color charged parton, but can only effectively resolve systems of adjacent partons. We therefore here consider a modified SCI model where the color exchange is instead made between the color string-fields. Thus, the single parameter is now instead giving the probability for a color octet exchange between pairs of string-pieces. Each such string-piece between two partons was formed from the color topology given by the preceding perturbative QCD processes. Obviously, it is only color exchange between the two overall color singlet systems depicted in Fig.~\ref{fig:ProtonColString} that will cause gap destruction through string formation over that central rapidity region. 

This gives a milder dependence on the increasing parton multiplicity and thereby a less strong reduction of the resulting gap rate at LHC energies. It is not necessary to derive the value of the new probability parameter from data. Instead, we have determined the exchange probability in this new model implementation by comparison to the original SCI model that described the Tevatron jet-gap-jet data. By matching the two model variants to give the same absolute amounts of gap events for $40<E_{T2}<50$ GeV, where they give very similar behavior, we obtain $P=0.9$ for the new model parameter.

\section{Model results and comparison with data}
\label{sec:Results}

The definition of a gap in jet-gap-jet events depends on the experimental setup. In accordance with the CMS study~\cite{CMSdata}, a jet-gap-jet event is here defined as having:
\begin{itemize}
\item At least two high-$p_\perp$ jets as defined by the cone algorithm with $\Delta R=\sqrt{\Delta\eta^2+\Delta\phi^2}=0.5$.
\item Two leading-$E_T$ jets with $|\eta|>1.5$, with one at positive and the other at negative pseudorapidity.
\item No charged particles with transverse momentum $p_T>0.2~\si{\giga \electronvolt}$ in the pseudorapidity region $|\eta|<1$.
\end{itemize}
The gap ratio is divided into bins depending on the rapidity separation $\Delta\eta$ of the two highest-$E_T$ jets and the transverse energy $E_{T2}$ of the second highest-$E_T$ jet. 

\begin{figure}[pht]
\centering
\begin{overpic}[width=0.8\textwidth,tics=10]{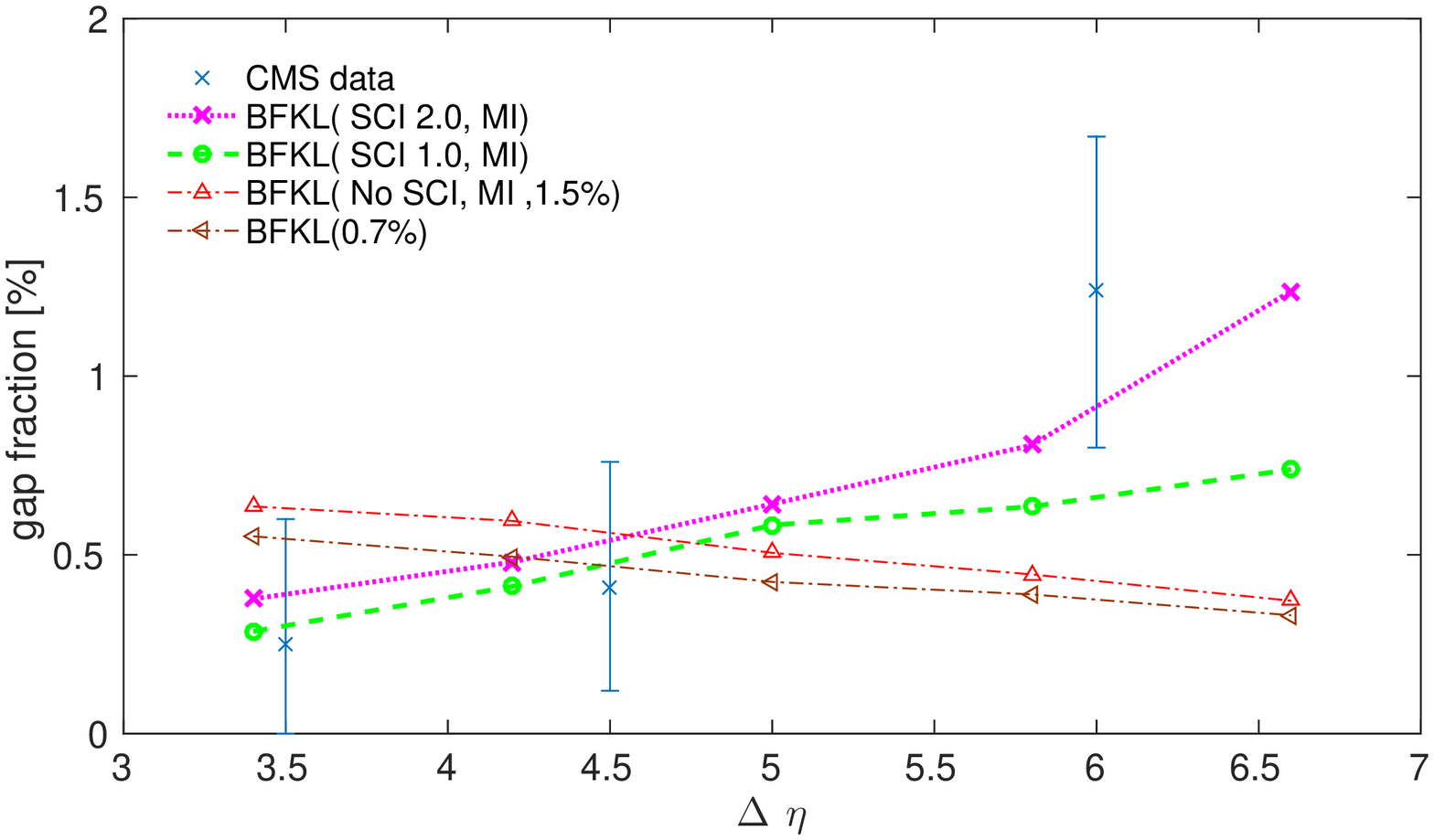}
 \put (40,44) {$40<E_{T2}<60~ \si{\giga \electronvolt}$}
\end{overpic}

\vspace*{-0.57cm}\begin{overpic}[width=0.8\textwidth,tics=10]{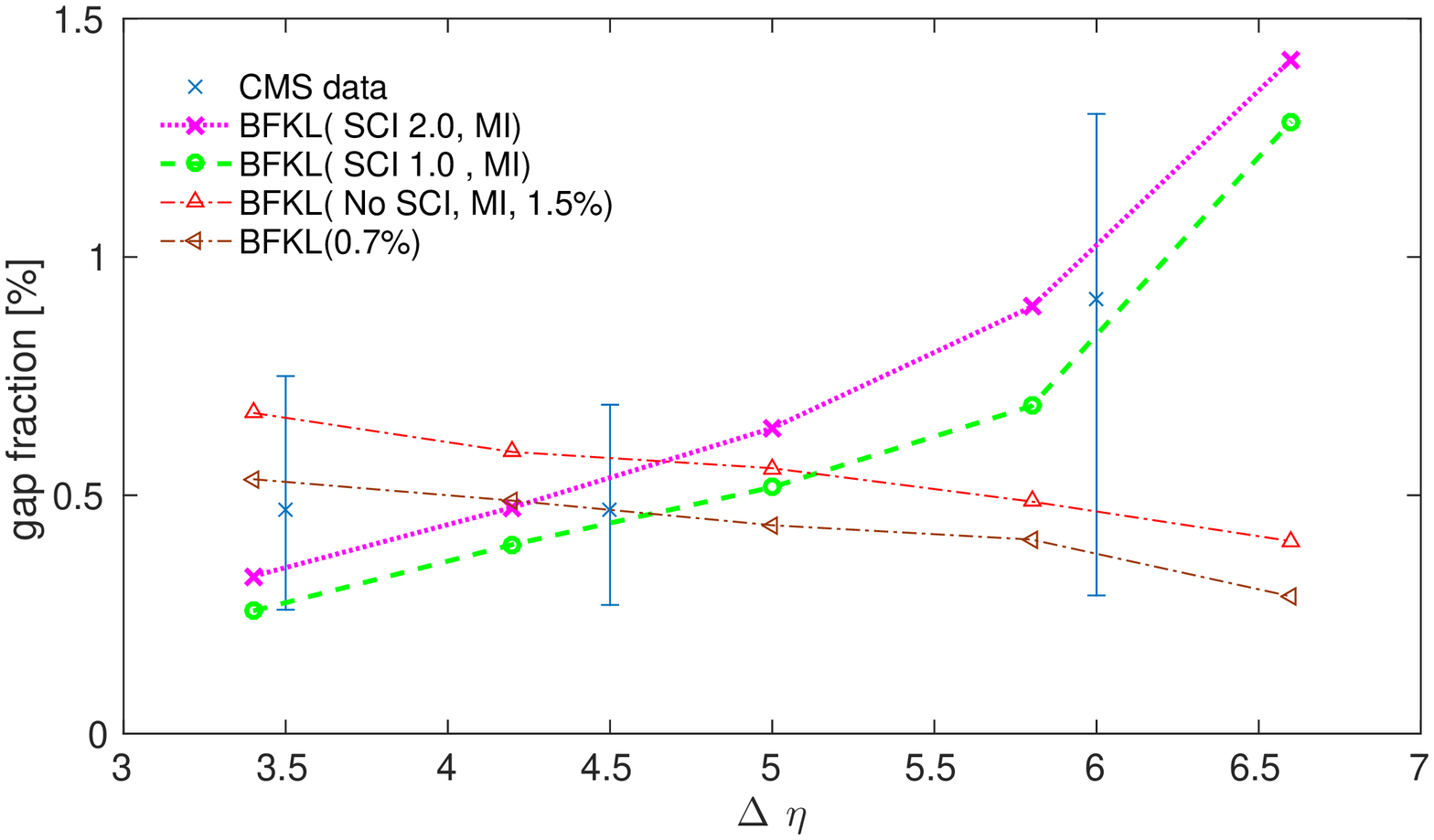}
\put (40,44) {$60<E_{T2}<100~ \si{\giga \electronvolt}$}
\end{overpic}

\vspace*{-0.57cm}\begin{overpic}[width=0.8\textwidth,tics=10]{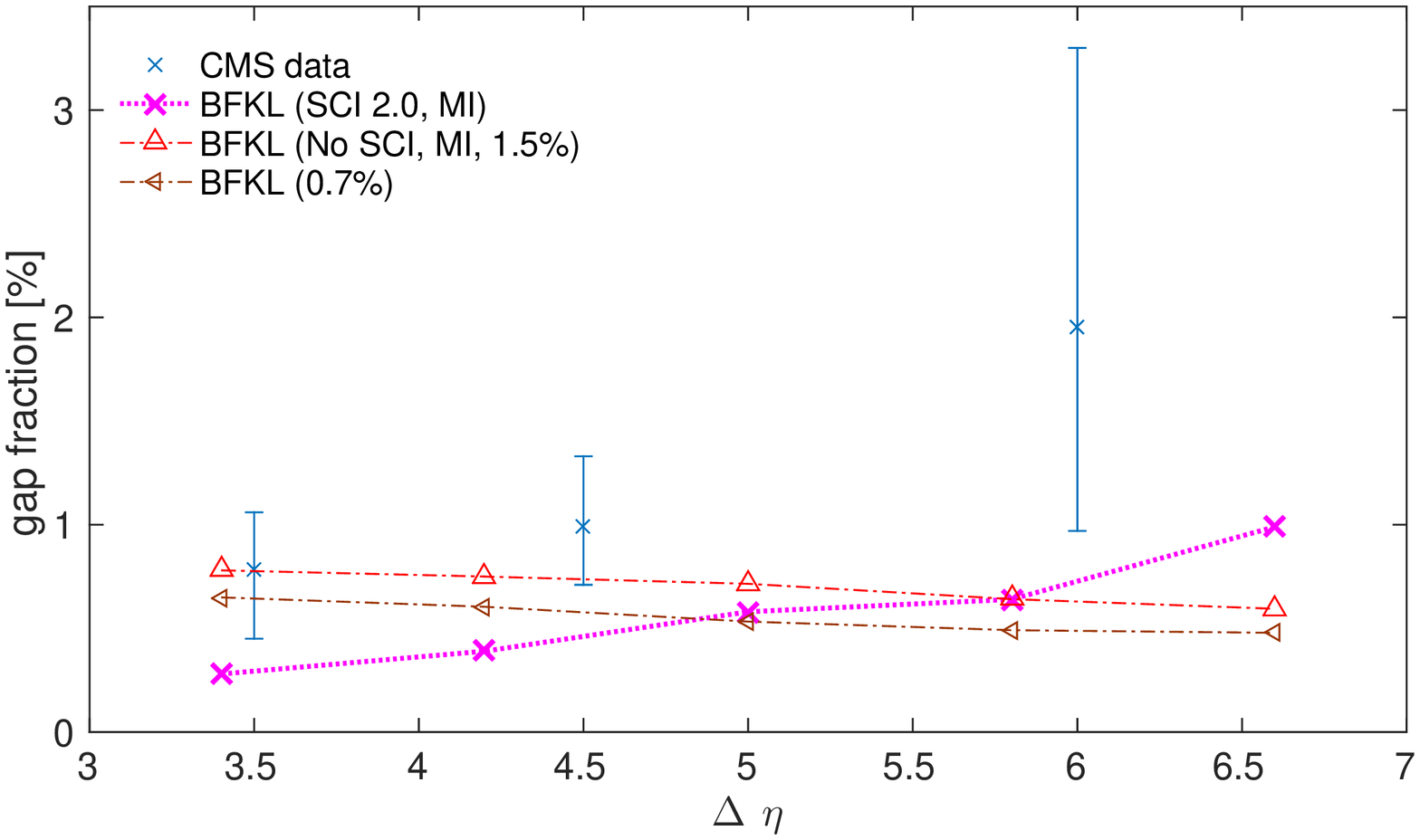}
\put (40,44) {$100<E_{T2}<200~ \si{\giga \electronvolt}$}
\end{overpic}
\caption[asd]{Ratio (in \%) of jet-gap-jet events over all 2-jet events versus the jet separation $\Delta\eta$ in pseudorapidity for different bins of jet transverse energy, $E_{T2}$, in pp-collisions at 7 TeV. CMS data (blue points) \cite{CMSdata} compared with the BFKL-model for color singlet two-gluon ladder exchange implemented in the Pythia~6.4 event generator with initial and final state parton showers, but with different additional gap destroying interactions: only overall gap survival factor of $S^2=0.7\%$ (brown dash-dotted curve), including multiple parton interaction (MI) and an overall gap survival factor $S^2=1.5\%$ (red dash-dotted curve), standard soft color interactions (SCI~v1, dashed green curve) and the modified version (SCI v2, magenta dotted curve).}
\label{fig:rapcms}
\end{figure}

Since we perform full Monte Carlo event simulations, we can apply the conditions on jets and gaps in detail. In particular the gap requirement can only be investigated properly when having a complete observable final state. For the inclusive two-jet event sample we use Pythia 6.4~\cite{Sjostrand:2006za} based on the dominating $2\to 2$ parton-parton hard scattering cross section in conventional leading order pQCD, and with the additional pQCD processes of initial and final state parton showers and multiple parton-parton scatterings, followed by Lund string model hadronization. Potential gap events are simulated using Pythia with our added implementations of the BFKL color singlet exchange and the SCI model, and resulting gap events selected. The fraction of jet-gap-jet events is then obtained as the ratio of these gap events divided by the inclusive two-jet event sample. 

\begin{figure}[ht]
\centering
   \begin{subfigure}[b]{1.0\textwidth}
   \includegraphics[width=1.0\linewidth,height=8cm,center]{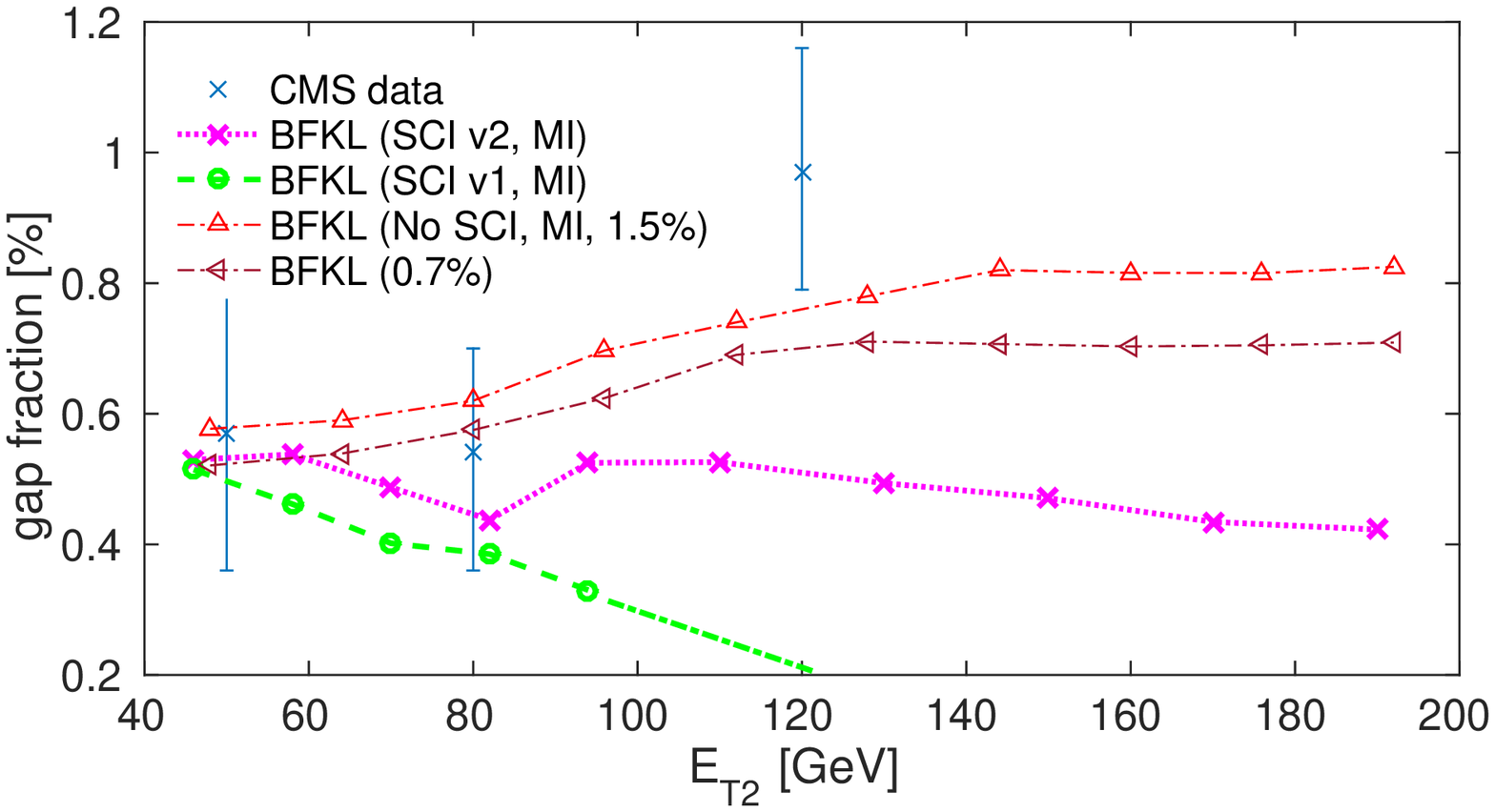}
   \label{fig:cmset} 
\end{subfigure}
\caption[asd]{Ratio (in \%) of jet-gap-jet events over all 2-jet events versus $E_{T2}$, the transverse energy of the second highest-$E_T$ jet, in pp-collisions at 7 TeV. CMS data (blue points) \cite{CMSdata} compared with the same BFKL-based models as in Fig.~\ref{fig:rapcms}.}
\label{fig:etcms}
\end{figure}

A characteristic feature of the BFKL equation is the relatively larger cross section for larger jet-$E_T$ compared to the leading order $2\to 2$ pQCD processes. As shown in Fig.~\ref{fig:rapcms}, a tendency of increasing jet-gap-jet ratio for larger $\Delta\eta$ is in fact observed both in the model and in the CMS data. Similarly, the basic BFKL cross section does not drop quite as fast with jet-$E_T$ as the standard processes and an increase of the gap fraction with $E_T$ is naively expected. This is also observed in the comparison of our model with the Tevatron data in Fig.~\ref{fig:Tevatron-ET} above, whereas the CMS data (Fig.\ \ref{fig:etcms}) has a statistically weaker hint of this.

The results on the gap fraction obtained from the model depends strongly on the gap destruction processes added to the BFKL exchange. With only the BFKL exchange process added to Pythia far too many gap events are produced. With only a simple gap survival probability one needs a value of about 15\% to fit the Tevatron data and only 1.5\% for the CMS data. If the standard multiple parton interactions (MI) are switched off in Pythisa, then the gap survival probability must be lowered to about 3\% to fit the Tevatron data and about 0.7\% for the CMS data. This shows that a non-trivial part of the gap destruction comes from additional multiple part-parton scatterings, which typically produce a color string-field across the central rapidity region.

One should note that the much stronger gap destruction at the LHC energy applies also when the $E_T$ and $\Delta\eta$ of the two-jet system is essentially the same. Thus an important fact seems to be the increased collision energy giving a larger phase space for additional scatterings and emissions. It is noteworthy that, for both Tevatron and LHC data, this constant gap survival, independent of the event dynamics, gives essentially the correct dependence in jet-$E_T$ but not in $\Delta\eta$.

Adding the previously successful SCI model to describe the gap destruction soft processes, one obtains a rather good description also of the CMS data on both jet separation $\Delta\eta$ and jet-$E_T$, except for the largest $E_T$-bin ($100<E_{T2}<200$~GeV) where a much too low rate of surviving gaps is noted. We note that this larger $E_T$ implies a larger maximum scale for both multiple parton scattering and parton shower processes that may cause gap destruction as discussed in the previous section. The modified SCI model gives an improved description of the CMS data. In particular, a much larger fraction of gap events at the largest jet-$E_T$ is obtained. This supports the physics basis for this model in terms of a limited resolution of soft color exchanges giving effectively fewer color charges and thereby fewer gap-destroying soft color exchanges. 

This simple model for soft color exchange to destroy potential gaps between jets is able to describe the data rather well. Nevertheless, it obviously needs to be developed and get a better theoretical foundation, e.g.\ along the lines discussed above in relation to QCD-based color exchange in diffractive deep inelastic scattering. More data is here essential for higher precision generally and, in particular, for large jet-$E_T$.

\section{Conclusions}
\label{sec:Conclusions}

The BFKL dynamics of QCD has received much attention theoretically, but experimentally accessible testing grounds are rare. This makes the observed jet-gap-jet phenomenon very interesting. It is, however, strongly affected by other QCD exchanges. Only a small fraction of potential gaps from the parton level color singlet two-gluon ladder exchange of BFKL survive to the final state of observable hadrons due to additional color exchanges and subsequent hadronization. It is therefore necessary to make very detailed investigations of these additional QCD processes both at perturbative and non-perturbative momentum transfer scales. 

We have here presented results from complete Monte Carlo event simulations where the basic hard parton-parton scattering via the BFKL exchange is complemented with the conventional treatments of additional parton-parton scatterings, initial and final state parton shower emissions and color string-field hadronization. This is found to produce a far too large rate of jet-gap-jet events as compared to the available data. The needed reduction can be parametrized as a simple-minded overall gap survival probability of 15\% at Tevatron energy and 1.5\% at LHC. 

More physics insights are obtained by adding additional color exchanges at a scale below the already included perturbative QCD processes, meaning in the region below 1-2 GeV but above the hadronization at the scale $\Lambda_{QCD}\sim 0.2$~GeV. Although this may seem a small interval, it is large on the logarithmic scale that is more relevant for many QCD processes, which then also have a larger coupling $\alpha_s$. 

We here apply the previously developed and phenomenologically successful soft color interaction model based on soft gluon, i.e.\ color octet, exchange with a fixed probability between pairs of partons emerging from the perturbative QCD processes. This could reproduce the earlier Tevatron data on the jet-gap-jet event rate without parameter adjustments. The new CMS jet-gap-jet rate can be reasonably well described in the jet-$E_T$ range overlapping with the Tevatron data, but at higher jet-$E_T$ much too few gaps result. This is due to the increased phase space for parton emission leading to increased parton multiplicity and a combinatorial growth of soft color exchanges. We have found that this problem can be cured by modifying the model based on the physical notion that soft gluon exchange cannot resolve all individual partons. We instead simulate the color exchange between string-field pieces with a fixed probability parameter. Adjusting this to agree with the previous model in the overlapping jet-$E_T$ range, we find a better description of the high-$E_T$ range of the CMS data. 

Although this simple model for soft color exchange is able to describe the jet-gap-jet data rather well, it needs to be developed and get a better theoretical foundation. Following the argument of limited resolution, we have suggested that it might be possible to apply the QCD-based dynamic color exchange formalism successfully applied in diffractive deep inelastic scattering. In any case, more jet-gap-jet data for in-depth QCD analysis would be very welcome.

\medskip
\noindent
{\bf Acknowledgments:}
We are grateful to Robert Ciesielski and the CMS team working on the jet-gap-jet analysis for interesting and helpful discussions. This work was supported by the Swedish Research Council under contract 621-2011-5107.


\end{document}